\documentstyle[12pt]{article}
\newcommand{\nl}{ {\hfill \break} }
\newcommand{\np}{ {\newpage } }

\newcommand{\N}{ \mbox{$\rm I\!N$} }
\def\R{\mbox{\rm {I\kern-.200em R}}}
\def\C{\mbox{\rm {I\kern-.520em C}}}

\newcommand{\vol}{ \mbox{\rm vol} }

\newcommand{\M}{ {\cal M} }

\newcommand{\beq}[1]{\begin{equation}\label{#1}  }
\newcommand{\eeq}{\end{equation}}
\newcommand{\bear}[1]{\begin{eqnarray}\label{#1} }
\newcommand{\ear}{\end{eqnarray}}

\begin{document}

\vspace*{-2.0cm}

\begin{tabbing}
\` {\small AIP 95-19}
  \\
\` {\small IPM-95-92}
  \\
\` {\small\it July 1995}
\end{tabbing}

\vskip 1.03cm

\centerline{DYNAMICS OF DIMENSIONS IN FACTOR SPACE COSMOLOGY}

\vskip 0.2cm

\centerline{U. Bleyer\footnote{E-mail: ubleyer@aip.de}\S*,
	    M. Mohazzab\footnote{E-mail: masoud@irearn.bitnet}\dag\ddag,
	    M. Rainer\footnote{E-mail: mrainer@aip.de
\\ \hspace*{0.5cm}
{\footnotesize\em
Financially supported by the DAAD and DFG grant Bl 365/1-1}
} \S  }


\begin{center}
{\S\
Gravitationsprojekt/Projektgruppe Kosmologie\\
Institut f\"ur Mathematik, Universit\"at Potsdam\\
P.O.Box 601553, D-14415 Potsdam, Germany
}
\end{center}

\vskip 0.1cm

\begin{center}
{*\
Urania Berlin e.V.\\
An der Urania 17, D-10787 Berlin, Germany
}
\end{center}

\vskip 0.1cm

\begin{center}
{\dag\
Institute for Studies in \\
Theoretical Physics and Mathematics\\
P.O.Box  5746, Tehran 19395, Iran
}
\end{center}


\begin{center}
{\ddag\
Physics Dept. Alzahra University \\
Tehran 19834, Iran
}
\end{center}


\vskip 0.22cm

\begin{abstract}

\vskip 0.14cm

We consider multidimensional cosmological models
with a generalized space-time manifold
$M= R\times M_1\times \cdots \times M_n$,
composed from a finite number of factor spaces $M_i$, $i=1,\ldots,n$.

While usually each factor space $M_i$ is considered
to be some Riemannian space of dimension $d_i\in \N$,
here it is, more generally, a fractal space, the dimension of which
is a smooth function of time $d_i(t)\in\R$.
Hence, besides the scale factor exponents ${\beta^i}=\ln a_i$
and their derivatives, we consider also the dimensions $d_i$
of the factor spaces as classical dynamical variables.

The classical equation of motions
and the corresponding Wheeler-de Witt equation
are set up generally, and the qualitative
behaviour of the system is discussed
for some specific model with $2$ factor spaces.
\end{abstract}
\vskip 4mm
\np

\section{Introduction}
\setcounter{equation}{0}
The observed correlation function of galaxy clusters and the fluctuation
of microwave background radiation seem to have fractal
structure \cite{LS}, \cite{COB}.
These kinds of observations suggest to attribute a fractal
structure to the universe. When the building blocks of
space-time or some of its subspaces have a fractal structure,
its dimension may have a noninteger value.

The assumption that space has a continuous dimension, was first proposed in
\cite{KMME} where a specific $(d+1)$-dimensional cosmological model with
 isotropic and homogenous $d$-dimensional spacelike slices
was proposed.
Its starting point is an extended Einstein-Hilbert Lagrangian in arbitrary
$d$ space dimension
together with a natural constraint between dimension
and scale factor of the universe.
The constraint arises when a cellular structure is attributed to space.

 Suppose the dimension of a space $M_0$ is $d_0$ and its size $a_0$.
This space can be constructed from a  finite number $N$ of
$d_0$-dimensional cells $e_0$.
 Now suppose we have a space $M$ of dimension $d > d_0$. In order to
 build such a space from the same number $N$ of cells, these
 should have an extra dimension $d-d_0$. Let us take the extra $(d-d_0)$-cell
 to be of  small size $\ell$ corresponding to some fundamental
length (e.g. the Planck scale).
 Then we get the $d$-dimensional volume of $M$ as
 $\vol_d(M) = N \vol_{d_0}(e_0)\ell^{d-d_0}$,
and analogously the volume of the $d_0$-space $M_0$ as
$\vol_{d_0}(M_0) = N \vol_{d_0}(e_0)$.
Sine the volume of a $d$-dimensional space of size $a$ is proportional to
$a^d$, the constraint
$$
 ({a \over \ell})^d=({a_0 \over \ell})^{d_0}
$$
arises. The model \cite{KMME} predicts that the universe quickly
becomes a FRW universe during its expansion.
Actually this universe may oscillate between a lower scale (the
fundamental length) and an upper scale
(the size of the universe now).
 During its expansion, the dimension of
the universe decreases to the present observed value while during
its contraction the dimension of the universe increases to a finite number.

 This model solves the horizon problem, since there is no
 starting time for the evolution of the universe, and therefore
 during several contractions and
expansions all points become correlated. Furthermore
there is no big bang singularity in the model.

On the other hand recent works \cite{BlRZ,Iv} on multidimensional cosmology,
generalizing the Kaluza-Klein idea,
use the possibility to assign further (however constant) dimensions $d_i$
of additional internal factor spaces $M_i$ to the universe
in its early stage of evolution.
Instead of a dynamical reduction of the spacial dimension (like in
\cite{KMME}) here the scales of the internal factor spaces contract.

Multidimensional geometric models
are an interesting class to study in cosmology, because
on one hand, they are rich enough to model features
of phenomenological interest, on the other hand they provide
a well defined minisuperspace. The latter is a convenient
starting point for covariant and conformally equivariant quantization,
with the energy constraint yielding the Wheeler-de Witt (WdW) equation.

Here, we generalize the above works by admitting the factor spaces
to have a fractal dimension which is a smooth function of time.
As an example we study the case
of two factor spaces, one flat and the other compact, where the latter
has a constant dimension.
In fact, the contribution from the scale factor of the compact space
with constant dimension is formally equivalent to some matter field
like the perfect fluid of \cite{KMME}.
Actually, we find it more conclusive (as compared to the standard approach)
to have all matter created from the geometry of space-time.

In Sec. 2 we give the setup of the canonical formalism for
a multidimensional cosmology with Riemannian and, more specifically,
constant curvature factor spaces.
After a proper reformulation
of the Lagrangian and Hamiltonian on the minisuperspace,
the canonical quantization can be applied.

Sec. 3 deals with conformally equivariant quantization on
a minisuperspace.
The first quantization of the energy constraint
is performed in a generally covariant and conformally equivariant
manner. Hence, there is no factor ordering problem.

Sec. 4 then considers the Lagrangian variation with dynamical dimensions,
where a constraint might be taken into account by a Lagrange multiplier.
However the resulting equations of motion
are in general too difficult to be solved analytically.

Therefore in Sec. 5 we consider another, very specific, Lagrangian model
and derive its equation of motion.

In Sec. 6 then the qualitative behaviour of this specific system
is discussed.

Sec. 7 refers to the WdW equation for this system and
Sec. 8 finally resumes the results.


\section{ Riemannian factor space cosmology}    
\setcounter{equation}{0}
A convenient reduction of the superspace of geometries is at hand
for the class of multidimensional geometries.
Usually,  with $d_i\in\N$, such a geometry is  described by a manifold
\bear{2.8}
M  =  \R\times M_1 \times\ldots\times M_n, \nonumber \\
D:= \dim M=1+d_1+\ldots+d_n                \nonumber \\
g\equiv ds^2  =  -e^{2\gamma} dt\otimes dt
     + \sum_{i=1}^{n} a_i^2 \, ds_i^2,
\ear
where    $ a_i=e^{\beta^i} $
is the scale factor of the factor space $M_i$ of dimension $d_i\in\N$,
Here we choose
$$
ds_i^2
=g^{(i)}_{{k}{l}}\,dx^{k}_{(i)} \otimes dx^{l}_{(i)}
$$
such that $ds_i^2$ is a regular bounded measure on $M_i$,
with a finite standard volume
$$
\vol_i:=\int_{M_i}ds_i <\infty.
$$
The scale factor exponents ${\beta^i}=\ln a_i$, $i=1,\ldots,n$,
provide a set of coordinates
for the $n$-dimensional minisuperspace $\cal M$ over $M$.
We subject the minisuperspace coordinates $\beta^1,\ldots,\beta^n$
to the principle of general covariance
w.r.t. minisuperspace coordinate transformations.

Like in \cite{BlRZ,Iv}, we restrict here
the $M_i$ to be Einstein spaces of constant curvature.
Then the Ricci scalar curvature of $M$ is
\bear{4.1}
R=e^{-2\gamma}\biggl\{
\biggl[ \sum_{i=1}^{n} (d_i \dot\beta^i) \biggr]^2
     + \sum_{i=1}^{n}
d_i[ { (\dot\beta^i)^2 - 2\dot\gamma\dot\beta^i + 2\ddot \beta^i } ]
\biggr\}
+\sum_{i=1}^{n} R^{(i)} e^{-2\beta^i}.
\ear
The action is usually taken in the standard form
\beq{4.2}
S=S_{EH}+S_{GH}+S_{M},
\eeq
where
$$
S_{EH}=\frac{1}{2\kappa}\int_{M}\sqrt{\vert g\vert} R\, dx
$$
is the Einstein-Hilbert action, $S_{GH}$
is the Gibbons-Hawking boundary term,
and $S_M$ some matter term.

Here we choose the boundary conditions such that
the terms with $\dot\gamma$, $\ddot\beta$ from (\ref{4.1}) and  $S_{GH}$
cancel out. Since we always have the possibility
to introduce one more dilatonic scale factor from the geometry
instead of some scalar matter field,
here we set $\delta S_M\equiv 0$ without restriction.

Then the variational principle of (\ref{4.2})
is  equivalent to a Lagrangian variational principle
over the minisuperspace $\cal M$, given in  coordinates $\beta^i$.
$$
S  =  \int L\, dt,
$$
\bear{4.3}
L  =  \frac{1}{2}{\mu}
\exp\Bigl\{-\gamma+\sum_{i=1}^{n}d_i\beta^i\Bigr\}
\biggl\{
\sum_{i=1}^{n}{d_i(\dot\beta^i)^2}
 -  \biggl[\sum_{i=1}^{n}{d_i\dot\beta^i}\biggr]^2 \biggr\} - V(\beta^i)
\ear
with
$$
V(\beta^i) =  {\mu}
\exp\Bigl\{\gamma+\sum_{i=1}^{n}d_i\beta^i\Bigr\}
\bigl[ - \frac{1}{2} \sum_{i=1}^{n} R^{(i)} e^{-2\beta^i}\bigr]
$$
where
\beq{4.4}
\mu:=\kappa^{-1}\prod_{i=1}^{n}\vol_i.
\eeq
Let us define a metric on $\M$, given in
coordinates  $\beta^i$, $i=1,\ldots,n$.
We set
\beq{4.5}
G_{kl}:=d_k \delta_{kl}-d_k d_l
\eeq
$k,l=1,\ldots,n$,
thus defining the tensor components
$G_{ij}$ of the minisuperspace metric
\beq{4.6}
G = G_{ij}d\beta^i\otimes d\beta^j.
\eeq

Then with a lapse function $N$, we obtain the Lagrangian
\beq{4.8}
L=\frac {\mu}{2N^2} G_{ij}\dot \beta ^i\dot \beta ^j
-V(\beta^i)
\eeq
with the energy constraint
\beq{4.9}
\frac {\mu}{2N^2} G_{ij}\dot \beta ^i\dot \beta ^j
+ V(\beta^i) = 0.
\eeq
A convenient gauge for $N$ is the harmonic one \cite{Iv,Rai} given by
\beq{4.7}
N^2:=\exp\biggl\{\gamma-\sum_{i=1}^{n}d_i\beta^i\biggr\} \stackrel{!}{=} 1.
\eeq
Nevertheless, here we do  not want to restrict to a specific gauge.
Unlike in \cite{BlRZ,Iv,Rai}, we will
prefer to implement a relation like
${\cal C}:=\gamma-\sum_{i=1}^{n}d_i\beta^i=0$ as constraint on the
configuration variables $\beta^i$, rather than as a gauge for $\gamma$.
Note that for a set of $m$ constraints ${\cal C}_k$, $k=1,\ldots,m$
we have to amend the potential
$V\beta^i$ by $-\sum_{k=1}^{m}\lambda_k {\cal C}_k(\beta^i)$, yielding
\beq{4.8c}
L=\frac {\mu}{2N^2} G_{ij}\dot \beta ^i\dot \beta ^j
-V(\beta^i)+\sum_{k=1}^{m}\lambda_k {\cal C}_k(\beta^i),
\eeq
with Lagrange multipliers $\lambda_k$.
Let us recall that, although in general the variation of (\ref{4.8c})
w.r.t. $\beta^i$ and $\lambda_k$
does not commute with the implementation of the constraints,
at least the set of solutions for  (\ref{4.8}) with
${\cal C}_k(\beta^i)=0$ resolved and substituted before the variation,
is a subset of the solutions of the variation of (\ref{4.8c}) including
the Lagrange multipliers.

Let us consider now the minisuperspace $\cal M$
Its signature of is Lorentzian for $n>1$ (see \cite{Iv}).
After diagonalization
of (\ref{4.5}) by a minisuperspace coordinate transformation
$\beta^i\to \alpha^i$ ($i=1,\ldots,n$), there is just one new
coordinate, say $\alpha^1$, in the direction of which the corresponding
eigenvalue of $G$ is negative. With a further (sign preserving)
coordinate rescaling, $G$ is equivalent to the
Minkowski metric \cite{Rai}. Hence $\cal M$ is flat.
Note that, unlike conformal flatness,
flatness is not an invariant property
under conformal transformation on $\cal M$.

In \cite{Rai} it was pointed out that
while $\beta^i\to\alpha^i$ is only a coordinate transformation
on $\M$, it transforms a multidimensional geometry
(\ref{2.8}) with scale exponents $\beta^i$ to another geometry
which is of the same multidimensional type (\ref{2.8}).
This has the same dimensions $d_i$ and first fundamental forms $ds^2_i$,
but new scale exponents $\alpha^i$ of the factor spaces
$M_i$. We can always perform the diagonalization of (\ref{4.5}) such
that $\alpha^1$ and hence $M_1$ belongs to the unique
negative eigenvalue of $G$. This $M_1$ is identified as ''external''
space. The scale factors of the ``internal''
spaces $M_2,\ldots,M_n$  contribute only
positive eigenvalues to the metric of $\M$.
$\alpha^1$ assumes in $\M$ the role played by time
in usual geometry and quantum mechanics.
In this way the ``external'' space
is distinguished against the ''internal'' ones, since its scale
factor provides a natural ''time'' coordinate on $\cal M$.
Note however that
the ``minisuperspace time'' $\alpha^1$ can be considered
as a time equivalent to $t$ in the underlying multidimensional geometry $g$
only if the space $M_1$ with $\alpha^1$ is
strictly expanding w.r.t. time $t$.
Then the Lorentzian structure of $\cal M$
provides  a natural ``arrow of time''
\cite{Ze}.

\section{Canonical minisuperspace  quantization}   
\setcounter{equation}{0}
Canonical quantization  essentially consists in
replacing the constraint equation (\ref{4.9}) by the WdW equation
\cite{ChZ}
\beq{5.1}
\left(-\frac{1}{2}\left[\Delta-\xi_c R\right]+V\right)\Psi = 0
\eeq
for a wave function  $\Psi$.

We set in the following
\beq{5.2}
N =: e^{-2f}
\eeq
and admit $f\in C^\infty(\cal M)$ to be an arbitrary smooth function on
$\cal M$.

In the time gauge given by $f$
the Lagrangian is
\beq{5.3}
L^f:= \frac{\mu}{2} {^{f}}\!G_{ij}(\beta)\dot{\beta}^i\dot{\beta}^j
- V^f(\beta )
\eeq
and the energy constraint is
\beq{5.4}
E^f:= \frac{\mu}{2} {^{f}}\!G_{ij}(\beta )\dot{\beta}^i\dot{\beta}^j
+ V^f(\beta ) = 0,
\eeq
where
\[
^{f}\!G = e^{2f}G \ \mbox{and} \ V^f = e^{-2f}V.
\]
With the canonical momenta
\beq{5.5}
\pi_{i} = \frac{\partial L^f}{\partial\dot{\beta}^{i}} =
\mu ^fG_{ij}\dot\beta^{j}
\eeq
this is equivalent to a Hamiltonian system given by
\beq{5.6}
H^f = \frac{1}{2\mu}(^f\!G)^{ij}\pi_i\pi_j + V^f
\eeq
and the energy constraint
\beq{5.7}
H^f = 0.
\eeq
The inverse of the minisuperspace metric is given by
$^{f}\!G^{-1} = e^{-2f} G^{-1}$, where for the system with Eq. (\ref{4.5})
the components of $G^{-1}$ are
\beq{5.8}
G^{ij} = \frac{\delta_{ij}}{d_i} + \frac{1}{1-\sum_{i=1}^{n}d_i}.
\eeq
At the quantum level $H^f$ has to be replaced by an operator
$\hat{H}^f$, acting in analogy to (\ref{5.7}) as
\beq{5.9}
\hat{H}^f\Psi^f = 0
\eeq
on wavefunctions, which are in
a conformal representation of weight $b$ given as
\beq{5.10}
\Psi^f=e^{bf}\Psi.
\eeq
Conformally equivariant quantization of $H^f$ from (\ref{5.6}) yields
$$
\hat{H}^f = e^{-2f}e^{bf}\hat{H}\ e^{-bf}
$$
\beq{WdWO}
\hat{H}^f = -\frac{1}{2\mu}\left[ \Delta^f -\xi_c R^f\right]
+{V}^f,
\eeq
on wave functions $\Psi^f = e^{bf}\Psi$,
where
\beq{cw}
		b=-(n-2)/2,
\eeq
\beq{6.14}
\xi_c=\frac{n-2}{4(n-1)},
\eeq
\beq{6.3}
\Delta^f=^f\!G^{ij}\nabla^f_i\nabla^f_j
\eeq
and both, $R^f$ and the covariant derivative $\nabla^f$, are determined
by the connection $\Gamma^f$ corresponding to the metric ${{}^f}G$.

The WdW equation (\ref{5.9}) is conformally equivariant
if and only if Eq. (\ref{5.9}) for any $f$ is equivalent to
\beq{5.13}
\hat{H}\Psi = 0
\eeq
where
\[
\hat{H} = \hat{H}^f\mid_{f=0}\ \mbox{and}\ \Psi =\Psi^f\mid_{f=0}
\]
are the Hamilton operator and the wave function in the gauge $f=0$.

\section {Variation with dynamical dimensions}
\setcounter{equation}{0}
We now pick up the Lagrangian (\ref{4.8c}) of Riemannian factor space
cosmology and consider it as Lagrangian with dynamical dimensions
$d_i$ of some fractal factor spaces.
Hence the dynamical configuration variables
are both $\beta^i$ and $d_i$ for $i=1,\ldots,n$.

In order to implement constraints ${\cal C}_i=0$
on the dynamics of dimensions, we
follow \cite{KMME}. Their constraint is given as follows:
Suppose each factor space is constructed from
a number $P$ of $d$-cells, each of which is a product of
a $d_0$-cell of macroscopic scale and a
$(d-d_0)$-cell of microscopic scale. From these cells we can
built a $d$ dimensional space. For finite $P$ and a vanishing measure on
the $(d-d_0)$-cells,
 the $d$-volume of the resulting space is zero. However if the
the $d-d_0$-cells have a length scale $\ell>0$ and nonvanishing
$(d-d_0)$-volume then the volume of each cell is $v_d=v_{d_0} {\ell}^{d-d_0}$.
Now if we write the scale of the factor space as $\ell e^{\beta}$,
we will have
$\ell^d e^{d\beta}= P v_d$ or $e^{d\beta}=e^{d_0\beta_0}$.
Since $d_0$ and $\beta_0$ are constant initial data for the
dynamics (take the present day values), we obtain the constraint
$d\beta=c$ with constant c.

 There are many alternatives for generalizing the above considerations
 to the case of multidimensional cosmology, e.g.:

1) $\sum_{i=1}^{n}\beta^i d_i= \gamma$

2) Constant $\sum_{i=1}^{m}\beta^i d_i= c_m$  for some $m<n$.

3) Constant $\beta^i d_i= c_i$ for $1\leq i\leq m\leq n$.

Slightly more general than case (1) is the following constraint:
\beq{constr}
\frac{\mu}{2}e^{-\gamma+\sum_{i=1}^{n}d_i\beta^i}={C},
\eeq
with a further function $\mu$, and
$C$ independent of the dynamical variables.

This constraint might also be reinterpreted  as a generalization
of a harmonic time gauge
of constant  $\mu$ and $\gamma$ as in \cite{Rai}.
With the harmonic time gauge  many interesting
cosmological models (see also \cite{BlRZ,Iv}) have been constructed.

Note that for constant curvature factor considered here we have
$R^{(i)}=K d_i(d_i-1)$.

Taking the constraint (\ref{constr}),
with  Lagrange multiplier $\lambda$,
the Lagrangian (\ref{4.8c}) is
\bear{}
L=\frac{\mu}{2}
\exp\Bigl\{-\gamma+\sum_{i=1}^{n}d_i\beta^i\Bigr\}
\bigl[ G_{ij}\dot \beta ^i\dot \beta ^j
+ K e^{2\gamma}\sum_{i=1}^{n} d_i(d_i-1) e^{-2\beta^i}\bigr]
\nonumber \\
+ \lambda \bigl(e^{-\gamma+\sum_{i=1}^{n}d_i\beta^i}-\frac{2C}{\mu}\bigr).
\ear

We assume that
\beq{volume}
\vol_i=m_i(d_i) {\tilde l_i}^{d_i},
\eeq
where $\tilde l_i$ is some characteristic length of $M_i$, and
$m_i$ is a function of $d_i$.
For dimension $d$ the coupling $\kappa$ is
\beq{kappa}
\kappa={\ell}^{d-1}
\eeq
for some fundamental length $\ell$.

Then with Eq. (\ref{4.4}) and
dimensionless $l_i:=\frac{\tilde l_i}{\ell}$
we obtain
\beq{mu}
{\mu}:=\ell\prod_{i=1}^{n} m_i(d_i) l_i^{d_i}.
\eeq
Variation w.r.t. $\lambda$ just reproduces the constraint.
Varying w.r.t. $d_k$ we obtain
$$
0{=}\frac{\partial L}{\partial d_k}
$$
$$
=C
\bigl\{
{\dot \beta^i} {\dot \beta^j} \bigl[
\frac{\partial G_{ij}}{\partial d_k}
+ (\frac{\frac{\partial m_k}{\partial d_k}}{m_k}+\ln l_k+\beta^k ) G_{ij}
\bigr] +
$$
$$
 K e^{2\gamma} \bigl[
\sum_{i=1}^{n} d_i(d_i-1) e^{-2\beta^i}
 (\frac{\frac{\partial m_k}{\partial d_k}}{m_k}+\ln l_k+\beta^k )+
(2 d_k-1) e^{-2\beta^k}
\bigr]\bigl\} +
$$
\bear{}
 \lambda
\{ \beta^ke^{-\gamma+\sum_{i=1}^{n}d_i\beta^i}
+\frac{2C}{\mu}
( \ln{l_k} +\frac{\frac{\partial m_k}{\partial d_k}}{m_k} ) \}.
\ear
The variation w.r.t. $\beta^k$ yields
$$
0{=}{d\over dt}\frac{\partial L}{\partial \dot\beta^k}
-\frac{\partial L}{\partial \beta^k}
$$
$$
=C
\bigl[
\dot G_{kj} \dot \beta ^j + G_{kj} \ddot\beta ^j
\bigr]
+\dot C G_{kj} \dot \beta ^j +
$$
\bear{}
K e^{2\gamma}
2C d_k(d_k-1) e^{-2\beta^k}
- \lambda d^k e^{-\gamma+\sum_{i=1}^{n}d_j\beta^j}.
\ear
For a general space $M_i$ the functions $m_i$ are too
complicated, and the equations above can hardly be solved  analytically.
Therefore we have taken the restriction to spaces of constant curvature,
and in the next section we consider a more specific model.

\section {Some specific Lagrangian model}
\setcounter{equation}{0}
Let us now consider more specific cases of the Lagrangian with
$ n $ factor spaces $M_i $ of dimension $ d_i $.
In what follows we specify the Lagrangian for the
cases that all factor spaces are of constant curvature.
Then, a space $M_i$ of positive curvature
is a $ d_i$-dimensional sphere.
For radius $r_i$, its volume is
\beq{spherevol}
\vol_i= {{2^{d_i} \pi^{d_i \over 2} }
		\over {(d_i -1) \Gamma({d_i\over 2})}}
 r^{d_i},
\eeq
where $\Gamma $ is the factorial function.
Hence here
\beq{sphere}
m_i(d_i)={ 1 \over {(d_i -1) \Gamma({d_i\over 2})}}
\qquad \mbox{and} \qquad
l_i= 2 \sqrt{\pi} r_i.
\eeq
In the case of an open factor
space, we can regularize the measure $ds_i$
such that the volume is
$\int ds_i = \vol_i < \infty $.
This could be done e.g. by a conformal
map reducing the radial extension $\infty\to l_i$.
For flat $M_i$ in Eq. (\ref{volume}) $m_i$ is constant.

In the following we assume:
\nl
a) One of the factor spaces, say $M_n$ is compact
with constant $R^{(n)}$,  all the other factor spaces $M_1,\ldots,M_{n-1}$
are flat.
\nl
b) The dimension $d_n$ of this space is
constant, all other dimensions are variable.
\nl
c) Here we choose $\gamma=\beta_0 d_1$, where $\beta_0=\beta_1(t_0)$
is the present value of the scale exponent of the external space $M_1$.

With Eqs. (\ref{mu}), (\ref{kappa}) and (\ref{sphere}),
normalizing all volumes with ${\tilde l}_i=\ell \qquad \forall i$,
choosing the constants $m_1,\ldots,m_{n-1}$ such that
$$
\prod_{i=1}^{n-1} m_i = {2\over \ell} (d_n-1) \Gamma({d_n\over 2}),
$$
the Lagrangian (\ref{4.8}) is
\bear{7.3}
L= e^{(\beta_1-\beta_0) d_1+\sum_{i=2}^{n} \beta^id_i}
\big\{ G_{ij} \dot\beta^i \dot\beta^j
+ R^{(n)} e^{2(\beta_0 d_1-\beta^n)}
\big\}
\ear
This is the generalization of \cite{KMME} to multidimensional
vacuum cosmology, here with  a potential from the  curvature
the compact factor space rather than the potential there.

In the following, we consider the alternative (3) of the previous section,
writing the constraints as
\beq{7.5}
{\beta^i d_i}=c_i
\eeq
for $1\leq i\leq m$.

While in the last section, in order to set up the equation of motion,
we had to variate the Lagrangian with respect to $\beta^i$, $d_i$, and
the Lagrange multiplier taking the constraint into account,
here we prefer to resolve the constraints first,
thus reducing the number of variables.

For simplicity we restrict to the case of $n=2$, $m=1$, $k=2$,
assuming that $M_1$ is a  flat space of variable dimension $d_1$
subject to a constraint (\ref{7.5}),
and $M_2$ a compact factor space with constant $d_2$.

Then the Lagrangian (\ref{7.3})
simplifies to
$$
L= e^{(\beta_1-\beta_0) d_1+\beta^2 d_2}
\big\{ [-(\dot\beta^1 d_1+ \dot\beta^2 d_2)^2
+(\dot\beta^1)^2 d_1+(\dot\beta^2)^2 d_2]
$$
\bear{Lagr}
+ R^{(2)} e^{2(\beta_0 d_1-\beta^2)}
\big\}
\ear
with
\beq{7.8}
 d_1={c\over\beta^1}
\eeq
where $c$ is a constant.
Its physical value can be found from the
observational value of the size of the universe \cite{KMME}.

In general, when the  constraints are implemented before
the variation of the Lagrangian,
we obtain a subset of the  full space of solutions.
Hence, unlike Sec. 4, here we resolve
the different constraints in the very beginning.

Then, besides (\ref{7.8}), the equations of motion will finally be
$$
\ddot\beta^1 +{\beta^1d_2\over {c-\beta^1}}\ddot\beta^2+
{1\over 2}(\dot\beta^1)^2({\beta_0 c\over {(\beta^1)^2}}
-{2c-\beta^1\over {c(c-\beta^1)}})
+ \dot\beta^1 \dot\beta^2 d_2 -
$$
\beq{7.7a}
{1\over 2}\dot\beta^2({d_2(d_2-1)\beta_0 \over {c-\beta^1}}
+{d_2^2\beta^1\over{c-\beta^1}})
+{\beta^1\beta_0\over{2(c-\beta^1)}} R^{(2)}
e^{2(\beta_0{c\over\beta^1}-\beta^2)}
=0
\eeq
and
$$
\ddot\beta^2 +{c\over {\beta^1(d_2-1)}} \ddot\beta^1
-{1\over {d_2-1}}(\dot\beta^2)^2+
(\dot\beta^1)^2({c\over{(\beta^1)^2(d_2-1)}}
+({bc\over{(\beta^1)^2}}+d_2){c\over {\beta^1(d_2-1)}})
$$
\beq{7.7b}
+ \dot\beta^2 {1\over {d_2-1}}-\dot\beta^1 {c^2\over{(\beta^1)^2(d_2-1)}}-
 ({1-{d_2\over 2}}) R^{(2)} e^{2(\beta_0 {c\over \beta^1}-\beta^2)}  =0
\eeq

\section {Qualitative behaviour of the system}
\setcounter{equation}{0}

{}From the Lagrangian (\ref{Lagr}) the Hamiltonian can be written

\beq{hamil}
H= e^{-\beta_0 d_1+\beta^2 d_2}G_{ij}\dot\beta^i \dot\beta^j -
R^{(2)} e^{\beta_0 d_1+(d_2-2)\beta^2}
\eeq
As the Lagrangian of this model is not an explicit function of time, the
Hamiltonian (\ref{hamil}) is a constant of motion.
When the system has a solution $E:=-{H\over {c}}$ is a positive constant.

Let us now first consider the case of constant $\beta^2:=q$.
At the early universe, it is $d_1\gg d_0:=d_1(t_0)$.
we  get the following equation
\beq{QU}
(\dot\beta^1)^2+{1\over {c^2}}(\beta^1)^2e^{2\beta_0 d_1-2q} R^{(2)}=
{E\over c^2}({\beta^1})^2 e^{\beta_0 d_1-q d_2}
\eeq
Eq. (\ref{QU}) is a $1$-dimensional mechanical system of constant
energy $E$.
The minimum value for $\beta^1$ (where $\dot \beta^1 =0$) is
$$
 {\beta^1}_{\min}={c\beta_0 \over {\ln({E\over R^{(2)}}})-q(d_2-2) }.
$$
Therefore a maximum value for the dimension of the early universe can
be derived
$$
{d_1}_{\max}={{ {\ln({E\over R^{(2)}}})-q(d_2-2) } \over \beta_0}
$$
On the other hand from \ref{QU}, one finds that
$$
\ddot \beta^1 > 0 \qquad \forall t
$$
Therefore $\beta^1_{min}$ is a local minimum. As a result the universe of
this model  contracts to a minimum and expands again. This behaviour
is similar to the previous work \cite{KMME}.
The behaviour of this universe at larger $\beta^1$
will be determined by the factor
$(\beta^1)^2e^{\beta_0 c\over {\beta^1}}$ in equation (6.2). Therefore
 with $k=\pm \sqrt{{E\over R^{(2)}}e^{-q(d_2-2)}-1}$,
$\beta^1$ will behave as $e^{kt}$.
Therefore this model predicts  the late expansion of the external space
to be inflationary.

In the case $\dot \beta^2\neq 0$,
for the early universe $\beta^1_{\min}$
is shifted to a larger value. Thus
the dynamics of the internal factor space $M_2$ prevents the external
factor space $M_1$ (our universe) from reaching a
singularity in the very early universe.
Another virtue of the second factor space to control the inflation.
The graceful exit of the
factor space $M_1$ can only be obtained as an effect of $\beta^2(t)$.

Complementary to the first  case above,
we have to examine also the situation for constant $\beta^1$.
There, the behaviour of $\beta^2$ can be solved
easily from (\ref{QU}). For $d_2=2$ we find $\beta^2$
proportional to  $\ln(t)$.

Therefore the scale factor of the second space decreases as ${1\over t}$.
As a result the general behaviour of the continuous dimensional space
is an exponential expansion and the second compact constant  dimensional
space contracts slowly. An exact  analysis of the behaviour , needs
the simulation of (\ref{7.7a}) and (\ref{7.7b}).

\section{WdW equation for the model}
\setcounter{equation}{0}
Recall that the
Lagrangian in an arbitrary gauge, with $N=e^{-2f}$, is given as
$$
L^f:=NL= {1\over N}G_{ij}\dot\beta^i\dot\beta^j-NV(\beta^i)
$$
\beq{Lf}
={1\over N}[G_{ij}\dot\beta^i\dot\beta^j-N^2V(\beta^i)].
\eeq
With the Lagrangian (\ref{Lagr}) and the
gauge $f={1\over 4}(\beta_0 d_1 -\sum_i \beta^i d_i)$
we get
\beq{Lag}
L^f=e^{{1\over 2}((\beta^1-\beta_0) d_1 + \beta^2 d_2)}
[G_{ij}\dot\beta^i\dot\beta^j
+ R^{(2)} e^{2(\beta_0 d_1-\beta^2)}].
\eeq
Now we consider the corresponding Hamiltonian
$$
H^f:=NH={1\over N}[G_{ij}\dot\beta^i\dot\beta^j+N^2V(\beta^i)],
$$
and change from the gauge
$f=-{1\over 4}(\beta_0 d_1 -\sum_i \beta^i d_i)$ to $f=0$.
Then, in the new gauge the Hamiltonian is
$$
H^0=G^{ij}\dot\beta^i\dot\beta^j+V(\beta^i)
$$
\beq{H0}
={1\over 4}G^{ij}\pi_i \pi_j+ V(\beta^i)
\eeq
where $V(\beta^i)=-R^{(2)}e^{2(\beta_0 d_1-\beta^2)}N^{-2}=
-R^{(2)}e^{\sum_i \beta^i d_i +\beta_0 d_1-2\beta^2}$.
Canonical quantization in this gauge yields
\beq{WdWop}
\hat H^0=-{1\over 4}\Delta +V,
\eeq
$$
\Delta=G^{ij}\nabla_i\nabla_j={1\over \sqrt{-\det G}}
\frac{\partial}{\partial \beta^i}
(\sqrt{-\det G}G^{ij}
\frac{\partial}{\partial \beta^j}).
$$
Recall that here $n=2$, hence Eq. (\ref{WdWO}) applies simply with
$\xi_c=0$ and $b=0$.
With $d_1={c\over \beta^1}$ and constant
$d_2$ the inverse of $G_{ij}$ is
\beq{invG}
(G^{ij})=
\left [\begin {array}{cc}
{\beta^1\over c}+{c \over c(1-d_2)-\beta^1} &
{c \over c(1-d_2)-\beta^1} \\\noalign{\medskip}
{c \over c(1-d_2)-\beta^1} &
{1\over d_2}+{c \over c(1-d_2)-\beta^1}
\end {array}\right
],
\eeq
and $\sqrt{-\det G}={1\over{\beta^1}}\sqrt{c(d_2-1)d_2\beta^1+c^2d_2} $.

Since with $n=2$ the conformal weight (\ref{cw}) is $b=0$,
a solution $\Psi=\Psi^0$ of the WdW equation $\hat{H}^0\Psi=0$
in the gauge $f=0$ is also a solution of the
the WdW equation $\hat{H}^f \Psi^f = e^{-2f}\hat{H}\Psi=0$
in the original gauge $f={1\over 4}((\beta_0 - \beta^1) d_1-\beta^2 d_2)$.

Actually here we have minisuperspace curvature $R[G]=0$ and
$\det G<0$ for $\beta_1>0$ and $d_2>1$.
Hence the minisuperspace $\cal M$ is
the $2$-dimensional flat Minkowski space,
like in the analogous case of constant dimensions.

In order to get a feeling for the qualitative structure of the WdW equation,
let us write it explicitly along a line of constant $\beta^2=q$.
There, at the limit of large $d_1$ it is
 \beq{WdW2}
 \biggl\{ (\frac{\partial}{\partial \beta^1})^2
+ {c\over {(\beta^1)^2}} \frac{\partial}{\partial \beta^1}
 + 4K d_2 (d_2-1)^2e^{c+q(d_2-2)}e^{\beta_0 c\over \beta^1} \biggr\}
\Psi\vert_{\beta^2=q}(\beta^1)=0
 \eeq
Orthogonally to this,  the WdW equation
along some line of constant  $\beta^1=\frac{c}{d_1}=p$ is
 \beq{WdW1}
 \biggl\{
\frac{d_1-1}{d_2 (d_1 d_2- d_1 +1)}   (\frac{\partial}{\partial \beta^2})^2
 - 4 Kd_2(d_2-1)e^{c+\beta_0 d_1}e^{\beta^2(d_2-2)}
\biggr\} \Psi\vert_{\beta^1=p}(\beta^2)=0.
 \eeq

Finally we want to compare our previous minisuperspace to the case
with $2$ constraints
$d_i\beta^i=c_i$, for {\em both}, $i=1$ {\em and} $i=2$.
There
\beq{invG2}
(G^{ij})= (\beta_1\beta_2-\beta_1 c_2-\beta_2 c_1)^{-1}
\left [\begin {array}{cc}
(\beta^2-c_2){\beta^1\over c_1} &
{\beta^1\beta^2} \\\noalign{\medskip}
{\beta^1\beta^2} &
(\beta^1-c_1){\beta^2\over c_2}
\end {array}\right
],
\eeq
and $\sqrt{-\det G}={1 \over\beta^1\beta^2}
\sqrt{c_1 c_2 (c_2\beta_1+c^1\beta^2-\beta^1\beta^2)} $.
Here, for $c_{1/2}>0$ and $d_{1/2}>1$, the minisuperspace
$\cal M$ is always Lorentzian. However, unlike the previous
example, it is no longer homogeneous, since
\beq{R2}
R[G]=-\frac{1}{2}\beta_1\beta_2/(c_2\beta_1+c^1\beta^2-\beta^1\beta^2)^2.
\eeq
For $\beta_1\to\infty$ or $\beta_2\to\infty$ the curvature decays
$R[G]\to 0$ and $\cal M$ becomes the usual homogeneous
Minkowskian space.
For $\beta_1\to\ 0$ or $\beta_2\to 0$ there appears a singularity
in the minisuperspace curvature, $R[G]\to \infty$ .
However, taking the conformal quantization scheme
seriously, we should be aware that the minisuperspace curvature
itself is not an invariant property of the quantum system
because it may be changed by conformal transformations.
More specifically, in our case the minisuperspace $\cal M$ is
$2$-dimensional, hence there exists a gauge $f$ such that
$\cal M$ is flat and therefore also homogeneous.
So for this model the inhomogeneity (\ref{R2})
is only a property for the specific gauge $f=0$.
According to  \cite{Rai},  in canonical  minisuperspace quantization
$\beta^1$ and $\beta^2$ are just coordinates for $\cal M$.
Hence the singularity of  (\ref{R2}) is the analogue
of a classical coordinate singularity.


\section{Conclusion}    
\setcounter{equation}{0}

We have investigated the effect of dynamical dimensions
of fractal factor spaces
on the evolution of a multidimensional cosmological model.
In a mathematically  closed approach, we could have
set up the differential geometry of fractal spaces in the
very beginning. This can essentially be done using a
definition of  generalized manifolds by simplicial complexes
(as exemplified e.g. in \cite{Hoef}).
However, for brevity,
here we rather preferred to derive the Lagrangian as for constant dimensions,
and  then to consider the dimensions as variables of just this Lagrangian.

More specifically, we discussed a multidimensional cosmological model with
two factor spaces, one of them flat with dynamical dimension, the
other compact with constant curvature and constant dimension.
In fact the latter behaves like a matter field.
By qualitative analysis, the behaviour of the system
shows that the universe, i.e. the factor space $M_1$,
contracts, passes through a state of minimum size
(maximum dimension), and expands.

For a static internal space (compare \cite{BlRZ}),
i.e. for constant $\beta^2=q$,
in the late expansion, for large $\beta^1$, the universe  inflates double
exponentially. Therefore, the dimension decreases very fast to its
minimum value.
Actually, the scale factor $\beta^2$ of the
compact constant dimensional space $M_2$ controls the
behaviour of the expansion of universe.
Eventually the dynamics of $\beta^2$ might be the only  way
to obtain a graceful exit to the effective model \cite{KMME}.
Again this model, as in the case of \cite{KMME}, has no big bang
singularity.
However, in order to find the complete behaviour of this
model, a more sophisticated analysis would be required.

In the case of a static internal space $M_2$,
near the maximun of $d_1$, i.e. minimum of $\beta^1$, the present model is
effectively represented by a model \cite{BlRZ,Rai} with only constant
dimensions. There the minimum value of $\beta^1$ could be related
to  the quantum creation  of the real Lorentzian space-time
{}from an Euclidian region. So, the present model is compatible with a
quantum creation of our universe.

We have further derived the WdW equation for the
minisuperspace of this model. Like in previous works on multidimensional
cosmology, here the metric describes a Minkowskian minisuperspace.
For the slightly more general case of both factor spaces
subject to the same type of constraint (\ref{7.5}), in the gauge
$f=0$, we find an inhomogeneity and singularities in
the minisuperspace curvature $R$. However, in the case of  only $2$
factor spaces, the conformal class of $G$ is in any case the flat
Minkowskian one.


\nl\nl
{\bf Acknowledgement}
\nl
{
Support by DFG grant Bl 365/1-1 and Schm 911/6-2
is gratefully acknowledged.
M. M.  thanks for hospitality at the Projektgruppe
Kosmologie of Universit\"at Potsdam.
}

%

\end{document}